\documentclass[12pt]{article}
\usepackage{wrapfig}
\usepackage{epsfig}
\usepackage{graphicx}
\begin{document}

\begin{center}
{\bfseries INVESTIGATION OF THE ANGULAR DEPENDENCE OF  
THE TENSOR ANALYZING POWER OF 9 GEV/C DEUTERON BREAKUP
\footnote{\it Talk given at the XVII-th International Baldin Seminar on
High Energy Physics Problems, 

~~ISHEPP XVII, 27 September- 2 October 2004, Dubna, Russia}} 

\vskip 5mm

L.S.~Azhgirey$^{1 \dag}$, S.V.~Afanasiev$^{1}$, V.V.~Arkhipov$^{1}$,
V.K.~Bondarev$^{1,2}$, Yu.T.~Borzounov$^{1}$, G.Filipov$^{1,3}$,
L.B.~Golovanov$^{1}$, A.Yu.~Isupov$^{1}$, A.A.~Kartamyshev$^{4}$,
V.A.~Kashirin$^{1}$, A.N.~Khrenov$^{1}$, V.I.~Kolesnikov$^{1}$,
V.A.~Kuznezov$^{1}$, V.P.~Ladygin$^{1}$, A.G.~Litvinenko$^{1}$,
S.G.~Reznikov$^{1}$, P.A.~Rukoyatkin$^{1}$, A.Yu.~Semenov$^{1}$, 
I.A.~Semenova$^{1}$, G.D.~Stoletov$^{1}$, A.P.~Tzvinev$^{1}$,
N.P.~Yudin$^{5}$, V.N.~Zhmyrov$^{1}$ and  L.S.~Zolin$^{1}$ 

\vskip 5mm

{\small  
(1) {\it Joint Institute for Nuclear Researches, 141980 Dubna, Russia} \\
(2) {\it St.-Petersburg State University, 198350 St.-Petersburg, Russia} \\
(3) {\it Institute of Nuclear Research and Nuclear Energy, 1784 Sofia, 
Bulgaria} \\
(4) {\it Russian Scientific Center "Kurchatov Institute", 123182 Moscow, 
Russia} \\
(5) {\it Moscow State University, 117234 Moscow, Russia} \\ 
$\dag$ {\it
E-mail: azhgirey@jinr.ru
}}
\end{center}

\vskip 5mm

\begin{center}
\begin{minipage}{150mm}
\centerline{\bf Abstract}
An angular dependence of the tensor analyzing power of the breakup 
of polarized deuterons at 9 GeV/$c$ has been investigated. 
The measurements have been made on hydrogen and carbon targets at 
angles in the range from 85 to 160 mr. The data  obtained are 
analyzed within the framework of the light-front dynamics using 
the deuteron wave functions for Paris and Bonn CD potentials, 
and the relativistic deuteron wave function by Karmanov et al. 
The experimental data are in rough agreement with calculations 
with the use of Karmanov's deuteron wave function. 
\end{minipage}
\end{center}

\vskip 5mm

{\large \bf 1. Introduction}

Investigations of polarization properties of the deuteron
fragmentation reaction  $(d,p)$  % [1-10] 
have amassed a convincing 
body of evidence that the description of the deuteron structure 
by means of wave functions derived from non-relativistic 
functions through the kinematical transformation
of variables  % \cite{reldwf} 
is liable to break down at  
short distances between nucleons. 
The main discrepancies between the expected and observed 
behaviour of data manifest themselves in the following facts.

({\bf i}) The expression for the tensor analyzing power $T_{20}$ 
of deuteron breakup, $A(d,p)X$, in the impulse approximation (IA) 
has the form  $T_{20} \sim w(k)[\sqrt{8} u(k) - w(k)]$,  where $u(k)$ 
and  $w(k)$ are the deuteron momentum space wave functions for $S$  
and $D$ states, respectively, and $k$ is the internal momentum 
of the nucleons in the deuteron (defined in the light-front system). 
With standard deuteron wave functions, the $T_{20}$ dependence on $k$ 
can be expected to change the sign at $k \sim 0.5$ GeV/$c$, but this 
expectation lacks support from experiment \cite{t20_data}. 

({\bf ii}) The recent measurements of the tensor analyzing
power $A_{yy}$ of the breakup of relativistic deuterons on nuclei at
non-zero angles of emitted protons \cite{azh5,large_t} show
that the measured $A_{yy}$-values at fixed value of the longitudinal
proton momentum have the dependence on the transverse proton
momentum that differs from that calculated with 
standard deuteron wave functions. 

({\bf iii}) The above-mentioned data show that the values of $A_{yy}$ 
being plotted at fixed values of $k$ tend to decrease as the 
variable ({\bf n k}) grows (vector {\bf n} is the unit normal 
to the surface of the light front). 

({\bf iv}) The pion-free deuteron breakup
process $dp\rightarrow ppn$ in the kinematical region close to that
of backward elastic $dp$ scattering at a given value of $k$ depends
on the incident momentum of deuteron \cite{azh3}. 

The foregoing forces one to suggest that an additional variable 
is required to depict the deuteron structure function at short 
distances, where relativistic effects are significant. 

     One approach that gives a relativistic description of the 
deuteron is light-front dynamics, one of several forms of 
Hamiltonian dynamics, first discussed by Dirac \cite{dirac}. 
Each form of dynamics is associated with a hypersurface on 
which the commutation relations for the generators of the 
Poincar\'e group are defined. There are many desirable 
features of the light-front dynamics \cite{fuda}. 

({\bf i}) High-energy experiments are naturally described using 
light-front coordinates: the fraction of the longitudinal 
momentum of the bound state taken away by the constituent 
particle in the infinite momentum frame is simply the ratio 
of the the plus momentum of the constituent to the total plus 
momentum of the bound state. 

({\bf ii}) There is a clean separation between center-of-momentum (internal) 
and relative (external) momentum variables in the light-front 
dynamics.

({\bf iii}) The vacuum for a theory with massive particles can be very 
simple on the light front: the vacuum with $p^+ = 0$ is empty, 
and diagrams that couple to this vacuum are zero. 

({\bf iv}) The generators of boosts in one, two, and plus directions are 
kinematic, meaning they are independent of the interaction.  
Thus, even when the Hamiltonian is truncated, the wave functions 
will transform correctly under boosts. 

These attractive features have led to the possibility that 
light-front dynamics is the best approach for calculating the 
spectrum and wave functions of relativistic composite system 
from an underlying field theory, such as quantum chromodynamics. 

      In previous papers \cite{azh1} the global features of proton
spectra in the region of transverse momenta of 0.5 -- 1 GeV/$c$,
produced in the reaction $(d,p)$ by unpolarized deuterons with an  
initial momentum of 9 GeV/$c$, were satisfactorily described on
the basis of the pole diagrams  within the framework of the
light-front dynamics. In those calculations the light-front
deuteron wave function was connected with the non-relativistic
deuteron wave function in a simple way, by the kinematical
transition from the equal-time variables to the light-front
variables. However, attempts to describe the tensor analyzing
power $A_{yy}$ of the reaction $^{12}C(d,p)X$ at an incident
deuteron momentum of 9 GeV/$c$ and a proton emission angle of 85
mr within the same approach have not met with the success
\cite{large_t}. The simple kinematical transition from a
non-relativistic deuteron wave function to the light-front one 
\cite{reldwf} presumably does not take into account essential 
features of the spin structure of a relativistic deuteron. 

     The relativistic deuteron wave function in the light front
dynamics was found in ref. \cite{karm1}. It is determined by six 
invariant functions $f_1,...,f_6$ instead of two ones in the 
non-relativistic case, each of them depending on two scalar 
variables $k$ and $z = cos ({\bf \widehat{k n}})$. 
The quantities ${\bf k}$ (the momentum of nucleons in the deuteron 
in their rest frame) and ${\bf n}$ (the unit normal to the light 
front surface) are defined by
\begin{equation}
x = \frac{E_p + p_{pl}}{E_d + p_d}, \quad
k = \sqrt{\frac{m_p^2 + {\bf p}_T^2}{4x(1-x)} - m_p^2}, \quad
({\bf n} \cdot {\bf k}) = (\frac{1}{2} - x) \cdot
\sqrt{\frac{m_p^2 + {\bf p}_T^2}{x(1-x)}}, 
\end{equation}
where $E_d$ and $p_d$ are the energy and the momentum of the
incoming deuteron, respectively, $p_{pl}$ is the longitudinal
component of ${\bf p}_1$, and $m_p$ is the mass of the nucleon.
It will be assumed further that ${\bf n}$ is directed opposite 
to the beam direction, i.e. ${\bf n} = (0,\ 0,\ -1)$.

     The expressions for the tensor analyzing power of the 
$(d,p)$  reaction using the above function are given
in ref. \cite{azhyud3}.

     Within the framework of this approach the following results 
on the description of the tensor analyzing power of the reaction 
$A(d,p)X$ have been obtained previously. 

({\bf i}) It was shown \cite{azhyud2} that calculations with 
Karmanov's deuteron wave function are in reasonably good 
agreement with the experimental data on the $T_{20}$ parameter 
of deuteron breakup on $H$ and $C$ targets with the emission of 
protons at 0$^\circ$ in the $k$ region from 0.4 to 0.8 GeV/$c$. 

({\bf ii}) A qualitative description of the momentum behaviour 
of the $A_{yy}$  parameter of the $^9Be(d,p)X$ reaction at a 
deuteron momentum of 4.5 GeV/$c$ and a detected proton angle of 
80 mr \cite{azh5} was obtained \cite{azhyud3}.

({\bf iii}) Rather good description of the $A_{yy}$ data for 
the $ ^{12}C(d,p)X$ reaction at 9 GeV/c and 85 mr \cite{large_t}
was  achieved \cite{azhyud3}. 

({\bf iv}) The experimental data on the tensor analyzing power 
$A_{yy}$ of the reaction $^9Be(d,p)X$ at an initial deuteron 
momentum of 5 GeV/$c$ and a proton emission angle of 178 mr are 
rather well reproduced with Karmanov's relativistic deuteron wave 
function as opposed to the calculations with the standard deuteron 
wave functions \cite{azh_pl}.

To get a more comprehensive picture of the angular behaviour 
of the tensor analyzing power of the deuteron breakup reaction, 
the $A_{yy}$ parameter in the interactions of polarized deuterons 
with hydrogen and carbon at 9 GeV/$c$ has been measured. 
The measurements have been made on hydrogen and carbon targets at 
angles in the range from 85 to 160 mr. 
\vskip 5mm

{\large \bf 2. Experiment}

     The measurements have been made at a polarized deuteron
beam of the JINR Synchrophasotron using the SPHERE setup
described elsewhere \cite{large_t}. 
     The polarized deuterons were produced by the ion source
POLARIS \cite{polaris}. 

     The tensor polarization of the beam was determined from the
asymmetry of protons with a momentum of $p_p \sim \frac{2}{3}p_d$
emitted at $0^\circ$ in the $A(d,p)X$ reaction \cite{zolin}, and it 
was $p_{zz}^+ = 0.798 \pm 0.002(stat) \pm 
0.040(syst)$ and $p_{zz}^-= -0.803 \pm 0.002(stat) \pm 0.040(syst)$ 
for positive and negative polarization directions, respectively.
     The vector polarization of the beam was monitored during the
experiment by measuring the asymmetry of quasi-elastic
$pp$-scattering on a thin $CH_2$ target placed in the beam
\cite{polarimeter}, and it values in 
different spin states were $p_{z}^+ = 0.231 \pm 0.014(stat) \pm 
0.012(syst)$ and $p_{z}^- = 0.242 \pm 0.014(stat) \pm 0.012(syst)$.

     A slowly extracted beam of tensor polarized 9-GeV/$c$ deuterons 
with an intensity of $\sim 5\cdot 10^8 \div 10^9$ particles per beam 
spill with a duration of 0.5 s fell on a liquid hydrogen target of 
30 cm length or on carbon targets with varied length.  
     The beam intensity was monitored by an ionization chamber. 
The beam positions and profiles at certain points of
the beam line were monitored by the control system of the accelerator
during each spill. The beam size at the target point was
$\sigma_x \sim 0.4$ cm and $\sigma_y \sim 0.9$ cm in the
horizontal and vertical directions, respectively.

     The data were obtained at secondary particle emission 
angles of 85, 130 and 160 mr (two measurements were made at 
115 and 145 mr), and secondary momenta between 4.5 and 9 GeV/c.
Along with the secondary protons, the apparatus detected the 
deuterons from inelastic scattering. The particles detected at 
given momentum were identified off-line on the basis of two 
independent time-of-flight (TOF) measurements with a base line 
of $\sim$ 34 m. The TOF resolution was better than 0.2 ns
($1\sigma$). Useful events were selected as the ones with two 
measured TOF values correlated. This allowed one to rule out 
the residual background completely. The values of the tensor 
analyzing power $A_{yy}$ obtained in the experiment are  
shown in Figs. 2, 3. The reported error bars 
are statistical only; possible systematic errors are estimated 
to be $\sim$ 5\%.

     The acceptance of the setup was determined by means of Monte 
Carlo simulation; the momentum and polar angle
acceptances were $\Delta p/p \sim \pm 2\%$ and $\pm
8$ mr, respectively. 
\vskip 5mm

{\large \bf 3. Formalism}

     The mechanism of the deuteron fragmentation $(d,p)$ can
be represented by the Feynman diagrams shown in Fig.~1. Here $d$
is the incoming deuteron, $p$ is the target proton, $p_1$ is the
detected proton, $b$ is the virtual (off-shell) nucleon, and $p_2,
p_3$ are nucleons. In addition to nucleons, one or more pions may
be produced at low vertices. Diagram (a) corresponds to the case 
where the detected proton results from deuteron stripping, and at 
the low vertex elastic $np$ scattering takes place. In diagrams (b)
and (c) the low vertices correspond to the charge exchange $np$ and
elastic $pp$ scatterings, respectively.

      The analyzing power $T_{\kappa q}$ of the $(d,p)$ reaction
is given by the expression
\begin{equation}
T_{\kappa q} = \frac{\int d\tau\,Sp\{{\cal M} \cdot t_{\kappa q}
\cdot {\cal M}^\dagger\}} {\int d\tau\,Sp\{{\cal M} \cdot {\cal
M}^\dagger\}},
\end{equation}
where $d\tau $ is the phase volume element, $\cal M$ is the
reaction amplitude, and the operator $t_{2q}$ is defined by
$$
<m\,|\,t_{\kappa q}\,|\,m^\prime> = (-1)^{1-m}
<1\,m\,1\,-m^\prime\,|\,\kappa\,q>,
$$
with the Clebsh-Gordan coefficients
$<1\,m\,1\,-m^\prime\,|\,\kappa\,q>$.

     The amplitude for the reaction $^1H(d,p)X$ in the light-front
dynamics is 
\begin{equation}
{\cal M}_a = \frac{{\cal M}(d \rightarrow p_1 b)} {(1-x)(M^2_d
- M^2(k))}{\cal M}(bp \rightarrow p_2 p_3),
\end{equation}
where ${\cal M}(d \rightarrow p_1 b)$ is the amplitude of the
deuteron breakup on a proton-spectator $p_1$ and an off-shell
particle $b$, and ${\cal M}(bp\rightarrow p_2 p_3)$ is
the amplitude of the reaction $bp\rightarrow p_2 p_3 $
(in the case of diagram (a), and with evident replacements of
indices for diagrams (b) and (c)).

%\vskip 5mm
\setlength{\unitlength}{0.8mm}   \thicklines
\begin{center}
\begin{picture}(150,40)
\put(15,0){({\bf a})}
\put(15,30){\circle{3}}
\put(23,14){\circle{3}}
\put( 0,30){\vector(1,0){13}}
\put( 0,32){$d$}
\put(17,31){\vector(3, 1){13}}
\put(32,32){$p_1$}
\put(16,28){\vector(1,-2){6}}
\put(21,22){$b$}
\put( 8,14){\vector(1,0){13}}
\put( 2,12){$p$}
\put(25,15){\vector(3, 1){13}}
\put(33,22){$p_2$}
\put(25,13){\vector(3,-1){13}}
\put(33, 4){$p_3$}
% \put(20, 0){{\bf  }}
\put(43,23){$-$}
\put(65,0){({\bf b})}
\put(65,30){\circle{3}}
\put(73,14){\circle{3}}
\put(50,30){\vector(1,0){13}}
\put(50,32){$d$}
\put(67,31){\vector(3, 1){13}}
\put(82,32){$p_2$}
\put(66,28){\vector(1,-2){6}}
\put(71,22){$b$}
\put(58,14){\vector(1,0){13}}
\put(52,12){$p$}
\put(75,15){\vector(3, 1){13}}
\put(83,22){$p_1$}
\put(75,13){\vector(3,-1){13}}
\put(83, 4){$p_3$}
% \put(70, 0){{\bf ¡}}
\put(115,0){({\bf c})}
\put(93,23){$+$}
\put(115,30){\circle{3}}
\put(123,14){\circle{3}}
\put(100,30){\vector(1,0){13}}
\put(100,32){$d$}
\put(117,31){\vector(3, 1){13}}
\put(132,32){$p_3$}
\put(116,28){\vector(1,-2){6}}
\put(121,22){$b$}
\put(108,14){\vector(1,0){13}}
\put(102,12){$p$}
\put(125,15){\vector(3, 1){13}}
\put(133,22){$p_1$}
\put(125,13){\vector(3,-1){13}}
\put(133, 4){$p_2$}
% \put(120, 0){{\bf ¢}}
\end{picture}
\end{center}
\normalsize
\vskip 3mm
Fig.~1. Feynman diagrams describing fragmentation of deuterons on protons.
\vskip 5mm

The ratio
\begin{equation}
\psi (x, p_{1T}) = \frac{{\cal M}(d \rightarrow p_1 b)} {M^2_d
- M^2(k)}
\end{equation}
is the wave function in the channel  $(b,\, N)$ 
given in \cite{karm1}; here
$p_{1T}$ is the  component of the momentum $p_1$ transverse to the 
$z$ axis. The light-front variables $p_T \equiv p_{1T}$ and $x$ 
(the fraction of the deuteron longitudinal momentum taken away by 
the proton in the infinite momentum frame) are given above.
The quantity $M^2(k)$ is given by
\begin{equation}
M^2(k) = \frac{m^2 + p^2_{1T}}{x} + \frac{b^2 + p^2_{1T}}{1-x},
\end{equation}
where $b^2$ is the four-momentum squared of the off-shell particle
$b$.

     The final expressions for the tensor analyzing power of
the $(d,p)$  reaction are given in ref. \cite{azhyud3}.
\vskip 5mm

{\large \bf 4. Results and discussion}

     It should be emphasized that the problem has no adjusted 
parameters. The invariant differential cross sections of processes  
taking place in the low vertices of the pole diagrams of Fig. 1, 
on the one hand, and the values of the invariant functions  
$f_1,...,f_6$  taken from ref. \cite{karm1}, on the other, were  
taken as input data. 
The contributions of the elastic and 
inelastic processes in the low vertex of the pole diagram were  
taken into account according to the parameterizations given 
in ref. \cite{azhrayu}. To account for the off-shell nature of 
particle $b$, the analytic continuations of the cross section 
parameterizations to the values of invariant variables 
$s^\prime = (b + p)^2, \quad t^\prime = (b - p_1)^2$  defined  
in the low vertex of the pole diagram at  $b^2 \neq m^2$  were  
used in the calculations. To obtain the values of functions 
$f_i(k,z)$ required for calculations, the spline-interpolation 
procedure between the table values given in ref. \cite{karm1} 
was used. 

     The experimental data on the tensor analyzing power $A_{yy}$ 
of the reactions $^1H(d,p)X$  and $^{12}C(d,p)X$ at an initial 
deuteron momentum of 9 GeV/$c$ and  proton emission angles of 85, 
130 and 160 mr are compared with the calculation results in Fig.~2. 
The data obtained on hydrogen and carbon targets are shown with 
empty and full circles, respectively. Above all it should be 
pointed out that data for both targets agree. By this is meant 
that nuclear targets are also appropriate to obtain information 
on the deuteron structure, as indicated before \cite{azh1}.
The previous data obtained at 9 GeV/$c$ on carbon at a proton 
emission angle of 85 mr \cite{large_t} are shown with full triangles. 

\begin{wrapfigure}{l}{9.0cm}
\mbox{\epsfig{figure=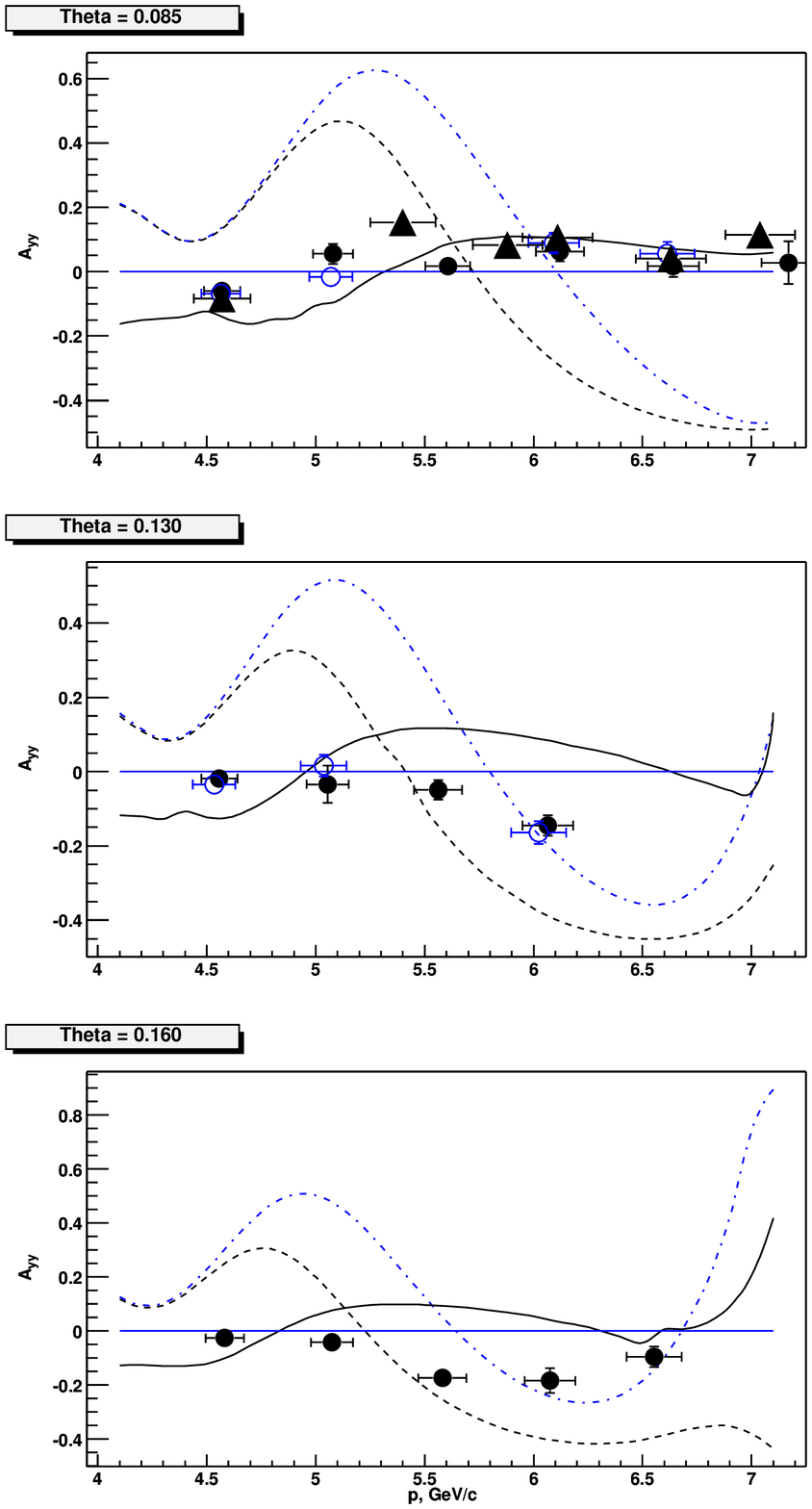,width=9.0cm,height=11.5cm}}
Fig.2. Parameter $A_{yy}$ of the reactions $^{1}H(d,p)X$ (empty circles) 
$^{12}C(d,p)X$ (full circles) at an initial deuteron momentum 
of 9 GeV/$c$ and  proton emission angles of 85, 130 and 160 mr 
as a function of the detected proton momentum. The data obtained 
in ref. \cite{large_t} are shown with full triangles. 
The calculations were made with the deuteron wave functions 
for the Paris \cite{paris} (dashed curves) and Bonn CD \cite{bonn} 
(dash-dotted curves) potentials. The solid curves were calculated 
with Karmanov's relativistic deuteron wave function \cite{karm1}.
\vspace{1mm}
\end{wrapfigure}

     It is seen that the experimental data at an angle of 85 mr 
are rather well reproduced with Karmanov's relativistic deuteron 
wave function as opposed to the calculations with the standard 
deuteron wave functions \cite{paris,bonn}. The data at angles 
of 130 and 160 mr are only in rough agreement with calculations 
using Karmanov's deuteron wave function, and, as before, they are 
in contradiction with calculations using standard deuteron wave 
functions. 

     The dependences of the analyzing power $A_{yy}$ on the 
transverse momentum of the protons $p_T$ at fixed values of 
total proton momenta $p$ close to 4.5, 5.0, 5.5, 6.0, 6.5, and 
7.0 GeV/$c$ are shown in Fig. 3.  These values of the total 
momentum of protons correspond to their longitudinal momentum 
fractions $x = 0.503, 0.558, 0.614, 0.670, 0.724,$ and 0.791.
It may be noted that the dependence of $A_{yy}$ on $p_T$ in the 
range of $p_T$ between 0.4 and 0.9 GeV/$c$ investigated in the 
present experiment is considerably more flat than in the range 
of $p_T$ from 0 to 0.6 GeV/$c$ as it was found in ref. \cite{azh5}. 
It is seen as well that $p_T$-behaviour of $A_{yy}$  counts in 
favour of Karmanov's relativistic deuteron wave function and 
definitely contradicts the predictions based on Paris and 
Bonn CD deuteron wave functions particularly at 
$p = 4.5, 5.0, 6.5,$ and 7.0 GeV/$c$.
\vskip 5mm

{\large \bf 5. Conclusion}

The results of this work can be summarized as follows. 

({\bf i}) The tensor analyzing power $A_{yy}$ of the reactions $^1H(d,p)X$  
and $^{12}C(d,p)X$ has been measured at an initial deuteron 
momentum of 9 GeV/$c$ and proton emission angles of 85, 115, 
130, 145 and 160 mr in the laboratory. The range of measurements 
corresponds to transverse proton momenta between 0.4 and 
0.9 GeV/$c$. 

({\bf ii}) The $A_{yy}$ data from the present experiment at 85 mr are in 
good agreement with the data obtained earlier \cite{large_t}. 

% \newpage
\begin{figure}[t]
\begin{minipage}[t]{0.48\linewidth}
\begin{center}
\includegraphics[width=9.0cm]{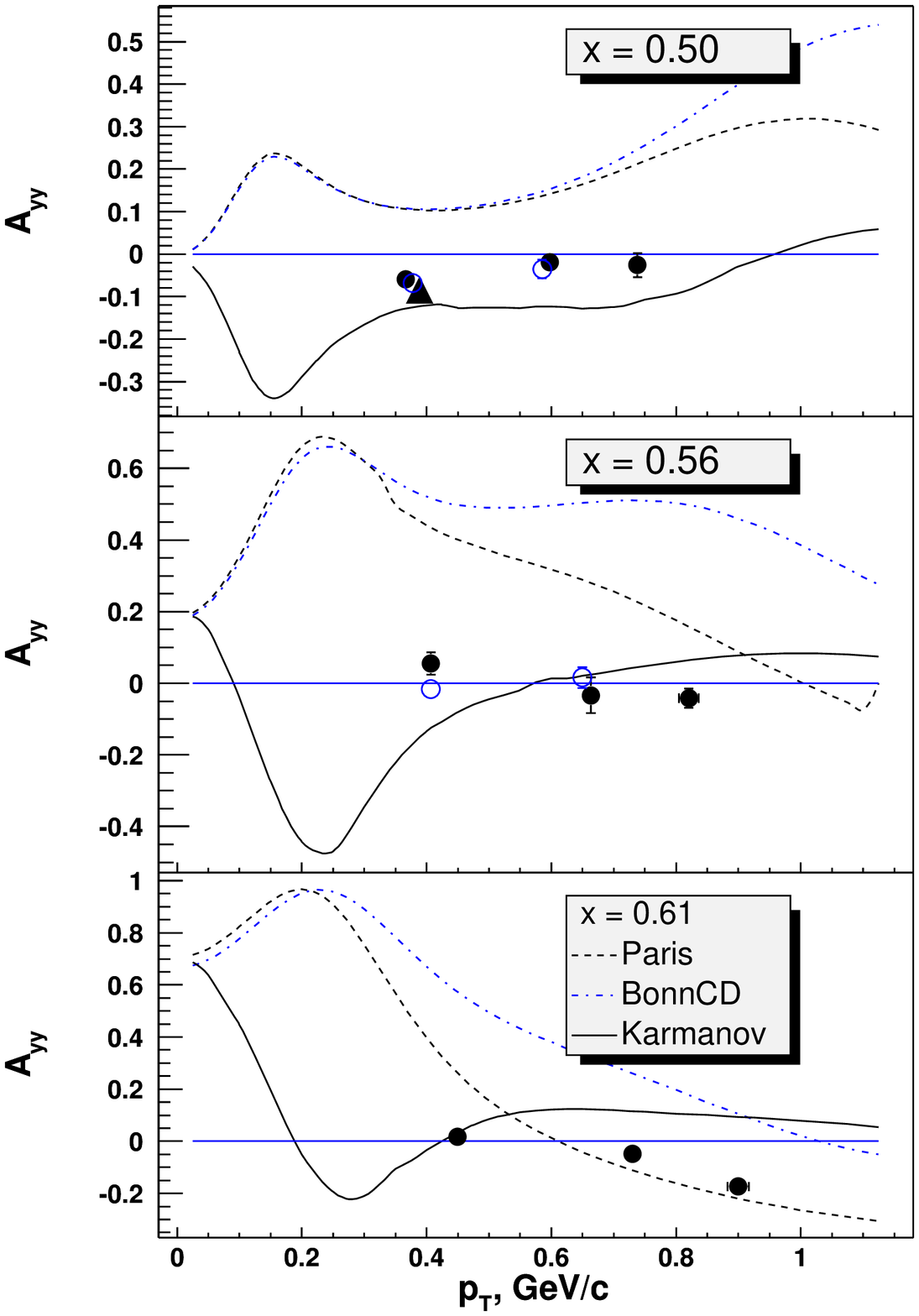}
\end{center}
\end{minipage}
% \hspace{0.04\textwidth}%
\begin{minipage}[t]{0.48\linewidth}
\begin{center}
\includegraphics[width=9.0cm]{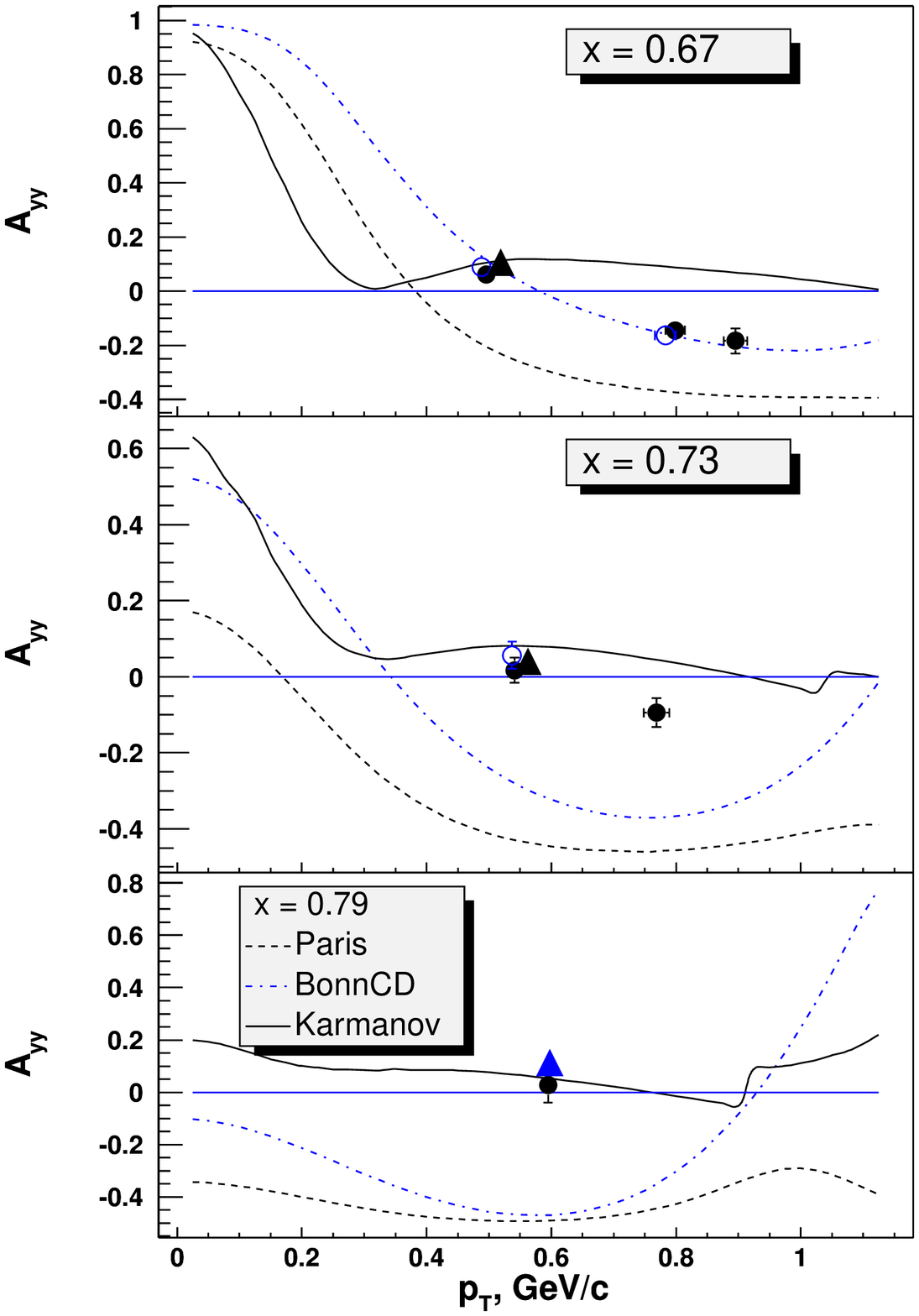}
\end{center}
\end{minipage}
\end{figure}
Fig. 3. Parameter $A_{yy}$ of the reactions $^{1}H(d,p)X$ (empty circles) 
$^{12}C(d,p)X$ (full circles) at an initial deuteron momentum 
of 9 GeV/$c$ and  proton emission momenta of 4.5, 5.0, 5.5  
(left panel) and 6.0, 6.5, 7.0 GeV/$c$ (right panel)
as a function of the transverse proton momentum $p_T$. 
The data obtained in ref. \cite{large_t} are shown with full triangles. 
The calculations were made with the deuteron wave functions 
for the Paris \cite{paris} (dashed curves) and Bonn CD \cite{bonn} 
(dash-dotted curves) potentials, and with the relativistic deuteron 
wave function \cite{karm1} (solid curves).
\vskip 5mm

({\bf iii}) The $A_{yy}$ data demonstrate an approximate independence 
on the A-value of the target, as it was pointed previously \cite{azh_pl}. 

({\bf iv}) The proton momentum dependences of $A_{yy}$ at the fixed 
values of proton emission angles and the transverse proton momentum 
dependences of $A_{yy}$ at the fixed values of longitudinal proton 
momentum fractions are in a better agreement with calculations using 
Karmanov's relativistic deuteron wave function instead of standard 
non-relativistic deuteron wave functions. While a quantitative 
description is not always achieved, the results obtained favours 
the description of the relativistic deuteron structure with 
a function depending on more than one variable. 
 
({\bf v}) Additional measurements of $A_{yy}$ and other polarization 
observables  at different initial deuteron momenta and 
various $p_T$ and $x$ are strongly desirable to provide 
the necessary experimental base to develop relativistic 
models describing the short-range structure of deuteron.

{\it Acknowledgment}. The authors 
are grateful to the LHE accelerator staff and the POLARIS 
team for providing good conditions for the experiment. 
They thank L.V.~Budkin, V.P.~Ershov, V.V.~Fimushkin, 
A.S.~Nikiforov, Yu.K.~Pilipenko, V.G.~Perevozchikov, 
E.V.~Ryzhov, A.I.~Shirokov, and O.A.~Titov for their 
assistance during the experiment. They are indebted to 
Profs. V.A.~Karmanov, V.I.~Komarov, and J.~Carbonell for 
helpful discussions. This work was supported by the Russian 
Foundation for Basic Research under grant No. 03-02-16224.


\begin{thebibliography}{99}
{\baselineskip 10pt
\bibitem{t20_data}
See, for example, V.G.~Ableev et al., Pis'ma Zh. Eksp. Teor. Fiz. 
{\bf 47}, 558 (1988); JINR Rapid Comm. {\bf 4[43]-90}, 5 (1990);  \\
V.~Punjabi et al., Phys. Rev. {\bf C39}, 608 (1989); \\
T.~Aono et al., Phys. Rev. Lett. {\bf 74}, 4997 (1995); \\ 
L.S.~Azhgirey et al., Phys. Lett. {\bf B387}, 37 (1996).
\bibitem{azh5}
V.P.~Ladygin et al., Few-Body Systems {\bf 32}, 127 (2002);
L.S.~Azhgirey et al., Yad. Fiz. {\bf 66}, 719 (2003).
\bibitem{large_t}
S.V.~Afanasiev et al., Phys. Lett. {\bf B434}, 21 (1998).
\bibitem{azh3}
L.S.~Azhgirey et al., Phys. Lett. {\bf B391}, 22 (1997); Yad. Fiz.
{\bf 61}, 494 (1998).
\bibitem{dirac}
P.A.M.~Dirac, Rev. Mod. Phys., {\bf 21}, 392 (1949).
\bibitem{fuda}
See, for example, M.G.~Fuda, Ann. Phys. (N.Y.) {\bf 197}, 265 (1990).
\bibitem{azh1}
L.S.~Azhgirey et al., Nucl. Phys. {\bf A528}, 621 (1991); Yad. Fiz.,
{\bf 46} 1134 (1987); Yad. Fiz. {\bf 53}, 1591 (1991).
\bibitem{reldwf}
See, for example, L.S.~Azhgirey and N.P.~Yudin, Yad. Fiz. {\bf 57}, 160 (1994);
C.E.~Carlson, J.R.~Hiller and R.J.~Holt, Ann. Rev. Nucl. Part. Sci.
{\bf 47}, 395 (1997).
\bibitem{karm1}
J.~Carbonell and V.A.~Karmanov, Nucl. Phys. {\bf A581}, 625 (1994).
\bibitem{azhyud3}
L.S.~Azhgirey and N.P.~Yudin, to be published in Yad. Fiz.;
see also Preprint arXiv:nucl-th/0311052 (2003).
\bibitem{azhyud2}
L.S.~Azhgirey and N.P.~Yudin, Preprint arXiv:nucl-th/0212033 (2002).
\bibitem{azh_pl}
L.S.~Azhgirey et al., Phys. Lett. {\bf B595}, 151 (2004).
\bibitem{polaris}
N.G.~Anishchenko et al., in {\it Proceedings of 5th International
Symposium on High Energy Spin Physics, Brookhaven, 1982)},
AIP Conf. Proc. {\bf 95}, 445 (1983).
\bibitem{zolin}
L.S.~Zolin et al., JINR Rapid Comm. {\bf 2[88]-98}, 27 (1998).
\bibitem{polarimeter}
L.S.~Azhgirey et al., PTE {\bf 1} 51 (1997) [Instr. and Exp. Tech.
{\bf 40}, 43 (1997)];
L.S.~Azhgirey et al., Nucl.Inst.and Meth. in Phys.Res. {\bf A497}, 340 (2003).
\bibitem{paris}
M.~Lacombe et al., Phys. Lett. {\bf B101}, 139 (1981).
\bibitem{bonn}
R.~Machleidt, Phys. Rev. {\bf C63}, 024001 (2001).
\bibitem{azhrayu}
L.S.~Azhgirey, S.V.~Razin and N.P.~Yudin, Yad. Fiz. {\bf 46} 1657 (1987).
}
\end{thebibliography}
\end{document}